\documentclass[12pt]{article}
\begin{document}

\begin{centering}
\Large{ An accurate description of quantum size effects in InP nanocrystallites
over a wide range of sizes}\\
\vspace*{1cm}
\large{Sameer Sapra, Ranjani Viswanatha and D. D. Sarma$^*$}\\
\vspace*{1cm} \textit{Solid State and Structural Chemistry Unit,
Indian Institute of Science, \mbox{Bangalore - 560012, India}}

\date{}
\end{centering}
\begin{abstract}
We obtain an effective parametrization of the bulk
electronic structure of InP within the Tight Binding scheme. Using
these parameters, we calculate the electronic structure of InP
clusters with the size ranging upto 7.5~nm. The calculated
variations in the electronic structure as a function of the
cluster size is found to be in excellent agreement with
experimental results over the entire range of sizes, establishing
the effectiveness and transferability of the obtained parameter strengths.
\end{abstract}
PACS No. 61.46.+w, 73.22.-f
\newpage

Effect of quantum confinement on electronic properties of
semiconductors has been a subject of intense study in the recent
years~\cite{revbrus} due to their potential applications in
various devices~\cite{alisci}. In the nanometer regime,
where the extensions of the electron and hole wave functions are
constrained by the particle size, the electronic and optical
properties are drastically different from the bulk material. For
example, the band gap of a semiconductor such as CdS can be varied
from its bulk value of 2.5~eV to about 4.5~eV for $\sim$~2~nm
sized clusters~\cite{dd}. This size quantization effect can be
qualitatively explained by the simplest quantum mechanical
particle-in-a-box problem where, upon decreasing the size of the
box, the energy separation between various levels increases due to
the confinement. Quantitatively, the effective mass
approximation (EMA)~\cite{effmass} describes such a size dependent
energy shift. The EMA, though suitable for larger clusters, fails
to correctly predict the band gap variation with the size for smaller
clusters, grossly overestimating the band gap of the small sized
semiconductor nanoparticles. Starting from the atomic limit, the
{\it ab initio} approaches, though accurate and hence desirable,
prove to be computationally expensive even for small sized
(~$\sim1.5$~nm) clusters~\cite{firstprinci}. An intermediate
approach that uses screened atomic pseudopotentials obtained from
self-consistent first principle calculations has been applied to
II-VI semiconductors~\cite{ramafrei} and recently to
InP~\cite{zunger1,zunger2}. The results are encouraging but this technique
too becomes computationally demanding for larger clusters; moreover,
such approaches are also somewhat empirical in nature in contrast
to the first principle calculations. Since
quantum confinement effects are appreciable for cluster sizes
comparable to the excitonic diameter in the bulk, which can be as
large as 15 nm in InP, it is obvious that an alternate methodology
is required to treat this problem. A semi-empirical tight-binding
(TB) scheme that borrows parameters from the bulk band structure
is less time consuming and has been applied with reasonable
success to II-VI~\cite{liplan}, III-V~\cite{s*calc} and group
IV~\cite{s*calc,niquet} semiconductors though the transferability
of bulk parameters to the nanometer regime is a controversial
issue. Our recent attempts~\cite{priya@prb,sameer} in this
direction suggest that some of the difficulties arise from an
improper or inaccurate tight binding (TB) parametrization of the
bulk band structures. It has been observed that the nearest neighbor
only model cannot describe the band structure accurately and that
inclusion of the next nearest neighbor interactions in the basis set is necessary
to obtain a good description of the bulk band dispersions.~\cite{sameer}
In the present letter, we show for InP that
a proper parametrization that includes the next nearest neighbor
interactions leads to an excellent description of experimental results
over the entire range of sizes.

Following our recent successful approach,~\cite{sameer}
we employ a TB method with the sp$^3$ orbital basis on In atom
and the sp$^3$d$^5$ orbital basis on the P atom. Along with the
nearest neighbor In-P interactions, our model also includes the
second nearest neighbor, In-In and P-P interactions.
The s$^*$
orbital used in the earlier TB models~\cite{liplan,s*calc,bassani}
is not required  in the present approach.

The TB Hamiltonian is given by
\begin{equation}
{\bf H} = \sum_{il} \epsilon_{il} a^{\dag}_{il} a_{il} + \sum_{ij}
\sum_{ll'} (t^{ll'}_{ij} a^{\dag}_{il} a_{jl'}  + {\rm h.c.})
\end{equation}
where $a^{\dag}_{il}$ and $a_{il}$ are respectively the creation
and annihilation operators for electrons at the atomic site, $i$
in the $l^{th}$ orbital. The onsite energy for the orbital $l$ at
the site $i$ is given by $\epsilon_{il}$.  The hopping interaction
strengths $t^{ll'}_{ij}$ depend on the type of orbital and
geometry of the lattice and are parametrized using the
Slater-Koster scheme~\cite{SK}.  In order to obtain the best
estimates for the onsite energies and the Slater-Koster
parameters, we obtained the band dispersions in InP along various
symmetry directions using the Linearized Muffin Tin Orbitals
within the Atomic Sphere Approximation (LMTO$-$ASA), after
converging the calculations with 43 $k$ points in the irreducible
part of the Brillouin Zone.~\cite{note1} The band gap obtained from the LMTO
method is 0.63~eV; an underestimation of the experimental value of 1.4~eV. Thus,
we correct the band gap to match the experimental value by shifting
the unoccupied bands by the appropriate amount. Then, we fit these {\it ab-initio}
band dispersions with those obtained from the TB model within a
least-squared-error approach by varying the TB parameter values.
Fig.~\ref{fit} shows the comparison between LMTO band dispersions
and the TB fitted band dispersions with optimized parameter values
(Table I), exhibiting an excellent agreement over the entire
valence and conduction bands.  It should be noted here that it is
not sufficient to get only the energies at various symmetry points
accurately described, as has been emphasized in the past. The effective
electron mass from our LMTO and TB calculations, deduced from the curvature
of the lowest conduction band, are 0.084 and 0.09, respectively,
showing an excellent fit of the curvature of the band dispersions near the $\Gamma$
point. They are also in good agreement with the reported
experimental values ranging between 0.068 to 0.084.~\cite{japrev}
In order to describe the evolution of the electronic structure of
clusters with the size, it is absolutely essential to obtain a
reliable description of the curvature of the band dispersion at
the extremal points.  The reliability of the present model and the
parameters obtained arises from an accurate description of both
the energies and the curvatures.~\cite{note2}

Similar to our previous studies of Mn-doped GaAs nanoparticles,~\cite{sapraNL02}
the clusters we generate consist of a central In atom surrounded
by the four nearest neighbor P atoms, followed progressively by
shells of In and P atoms, respectively.  The effective diameter of
the cluster is calculated assuming that the particles are
spherical in shape using the formula
\begin{equation}
d = a~[\frac{3N_{at}}{4{\pi}}]~^{1/3}
\end{equation}
where $a$ is the bulk lattice parameter (5.861~\AA~ for
InP~\cite{atompara}) and $N_{at}$ is the number of atoms in the
cluster. In our calculations, the largest cluster has $N_{at}$ =
9527 atoms and a diameter of 77.1 \AA~ containing 60328 orbitals.
We passivate the clusters with hydrogen atoms at the outermost
layer to remove the dangling bonds and obtain the eigen-value
spectrum for clusters with different sizes using the Lanczos
algorithm~\cite{lanc}.

Variations of energies of the top of the valence band (TVB) and
the bottom of the conduction band (BCB) as a function of cluster
diameter are shown in Fig.~\ref{tvbbcb}. The results clearly show
the systematic shifts of both TVB and BCB away from those of the
bulk, marked by dashed lines, with decreasing cluster size as a
consequence of quantum confinement. The positions of TVB and BCB
for four specific cluster sizes have been reported in Ref.~\cite{zunger2}; we
show these results in the same figure indicating a good
agreement for the common range of sizes.~\cite{note4} In order to compare the
present results with the experimental estimates of band edges
obtained from the optical absorption experiments, it is necessary
to take into account the excitonic binding energies at different
cluster sizes; we use the form~\cite{kama} $3.572 * e/\epsilon d$
for the excitonic binding energy, where $d$ is the diameter of the
cluster and $\epsilon$ is the the dielectric constant, $12.4$ for
InP. Thus derived calculated band-gap variation with cluster size
is shown in the inset in Fig.~3, exhibiting a systematic and
pronounced variation. The dependence of the band gap on the
diameter of the cluster is expected to be given by an inverse
square, $1/d^2$, law according to EMA. However, such a
relationship is completely incompatible with the present results.
We find that $1/d^x$ with $x=1.04$ provides the best fit of the
calculated results, as shown by the dotted line in the inset,
suggesting an approximately $1/d$ dependence, instead of the
$1/d^2$ one. However, this fit is not satisfactory, particularly
for large cluster sizes (see inset, Fig.~\ref{comp}). We find that the
expression, $100*(5.8d^2 + 27.2d + 10.4)^{-1}$, provides a good
description of the band gap variation with the cluster diameter
over the entire range of sizes, as shown by the solid line in the
inset.

The variation of the band gap, $\Delta E_g$, with respect to the
bulk band gap is compared with experimental values and with
results obtained from other calculations in the main frame of
Fig.~\ref{comp}. The excellent agreement of the present calculated
results with the experimental
data~\cite{micicali97,micicali96@apl,micicali96@jpc,micicali98}
%can be observed
over the entire range of sizes is evident from this comparison.
The expected band gap variation on the basis of EMA~\cite{effmassval}, shown in the
figure with a thin line, is in gross disagreement with the
experimental results. Though the parameterized pseudopotential
result~\cite{zunger2} qualitatively predicts the variation of band
gap accurately and tends to agree with the experimental band gap
as one approaches the atomic limit and the bulk limit, it slightly
underestimates the band gap in the nanometer regime. The
sp$^3$d$^5$s$^*$ parametrization scheme that was recently
reported~\cite{s*calc} overestimates the band gap systematically
over the entire range of sizes~\cite{note3}.

In conclusion, we have shown that an effective
parametrization within the tight binding model allows us to
describe the bulk electronic structure of InP without the
need to invoke the fictitious s$^*$ orbital. We have also established
the transferability of these parameters to obtain electronic
structures of clusters over a wide range of sizes; thus calculated
variation in the band gap with the cluster size is in excellent
agreement with experimental results. Taking into account the
accuracy of this method, the ease of implementation and the
computational cost, it should
be possible to apply this method to perform molecular dynamics of
large clusters.

We thank O. K. Aderssen and O. Jepsen for the LMTO codes,
P. Mahadevan for the LAPW calculations and
the Department of Science and Technology for funding the
project.

\newpage
\begin{figure}[h]
\caption{\label{fit} Band structure obtained from LMTO and fitted
using TB parametrization for bulk InP. The zero of the energy
corresponds to the top of the valence band.}
\end{figure}
\begin{figure}[h]
\caption{\label{tvbbcb} Variation of top of the valence band (TVB) and
bottom of the conduction band (BCB) with effective
diameter of the clusters. The pseudopotential results from Ref.~\cite{zunger2}
are also shown. The dashed lines show the TVB and the BCB for
bulk InP.}
\end{figure}
\begin{figure}[h]
\caption{\label{comp} Comparison of calculated band gap variation
with experimental data and other theoretical results. The inset
shows the $100*(5.8d^2 + 27.2d + 10.4)^{-1}$ fit (solid line) and the
$1/d^{1.04}$ fit (dotted line) to the present calculations.}
\end{figure}

\newpage
\center{Table~I}\\
%\flushleft
\center{TB parameters (in eV) obtained from least squared error fit to
LMTO band dispersions for InP and parameters for hydrogen passivation.}

\begin{centering}
{
\vspace*{0.5cm}
Onsite Energies
\vspace*{0.5cm}

\begin{tabular}{|l|ccc|}
\hline
& ~s~ & ~p~ &  ~d~ \\
\hline
  ~In~   & ~$-$1.53~ & ~3.92~ & \\
  ~P~   & ~$-$10.24~ & ~$-$0.63~ & ~16.62~\\
  ~H~   & ~$-$0.7412~& &\\
\hline
\end{tabular}

\vspace*{1cm}Slater Koster Parameters\\
\vspace{0.5cm}
\begin{tabular}{|l|cccccccc|}
\hline & ~ss$\sigma$~ & ~sp$\sigma$~ & ~sd$\sigma$~
& ~ps$\sigma$~ & ~pp$\sigma$~ & ~pp$\pi$~  & ~pd$\sigma$~ & ~pd$\pi$~ \\
\hline  ~In-P~  & ~$-$1.43~ & ~2.19~ & ~$-$2.72~ &
~$-$1.63~ & ~3.35~ & ~$-$0.66~ & ~$-$3.38~ & ~3.35~ \\
~In-H~  & ~$-$2.944~ & ~2.76~ & ~$-$1.36~ & \\
\hline
\end{tabular}

\vspace{0.5cm}
\begin{tabular}{|l|cccc|}
\hline & ~ss$\sigma$~ & ~sp$\sigma$~ & ~pp$\sigma$~ & ~pp$\pi$~ \\
\hline ~In-In~  & ~$-$0.21~ & ~0.00~ & ~0.14~ & ~$-$0.01~\\
 ~P-P~  & ~$-$0.01~ & ~0.14~ & ~0.70~ & ~$-$0.02~\\
\hline
\end{tabular}

}
\end{centering}


\begin{thebibliography}{99}
\bibitem{note1} We compared our LMTO results with those
obtained using the linearized augmented plane-wave (LAPW) method
and found them to be in excellent agreement.
\bibitem{note2} At this point, we would like to emphasize upon the accuracy
of the present calculations in comparison with the sp$^3$d$^5$s$^*$ results
from ref~\cite{bassani}. The total root mean square error in the TB fit at the
symmetry points listed in Table V, ref.~\cite{bassani} is 0.354~eV for the
sp$^3$d$^5$s$^*$ model and 0.234~eV for our model.
\bibitem{effmassval}
The effective mass of electron for InP is 0.08m$_0$ and that of
the hole is 0.61m$_0$ where m$_0$ is the mass of the free electron
(taken from Ref.~\cite{zunger1}).
\bibitem{note3} The Coulomb interaction term has been subtracted from the
calculations of Ref.~\cite{s*calc} which uses the $sp^3d^5s^*$ orbital basis
in the TB model.
\bibitem{note4} The results of Ref.~\cite{zunger2} were rigidly shifted in order
to ensure that the zero of the energy scale coincided with the top of the valence
band in order to be consistent with the present convention.
\end{thebibliography}
\end{document}